# Cryogenic-specific reddish coloration by cryoplasma: New explanation to color diversity of outer solar system objects

Noritaka Sakakibara[1][2], Phua Yu Yu[1], Tsuyohito Ito[1], Kazuo Terashima[1][3]

Running head: color diversity of outer solar system objects

[1] Department of Advanced Materials Science, Graduate School of Frontier Sciences, The University of Tokyo, 5-1-5 Kashiwanoha, Kashiwa, Chiba 277-8561, Japan

[2] n.sakakibara@plasma.k.u-tokyo.ac.jp

[3] Corresponding author: kazuo@plasma.k.u-tokyo.ac.jp

# ABTRACT

Reddish coloration and color diversity among icy bodies in the outer solar system are one of the significant clues for understanding the status and history of the solar system. However, the origin of color distribution remains debatable. Here, we demonstrate reddish coloration that is stable only at cryogenic temperatures in a laboratory experiment. The reddish coloration was produced on methanol- and water-containing ice irradiated with nitrogen-containing cryoplasma at 85 K. The reddish color visually faded and disappeared at 120–150 K as the ice was heated, unlike well-known refractory organic tholins that are stable even when heated to room temperature. This temperature dependence of reddish coloration under cryogenic conditions could provide a new possible explanation for the absence of ultra-red coloration closer to the Sun in the outer solar system. Our result implies that a reddish material specific to cryogenic environments is useful for the investigation of color diversity and formation mechanism of the outer solar system.

# 1. INTRODUCTION

Reddish coloration and its diversity are one of the significant characteristics widely observed in the outer solar system objects at larger heliocentric distances (Jewitt 2002b; Sheppard 2010; Stern et al. 2015; Weaver et al. 2016; Stern et al. 2019). Ultra-red objects occur only relatively far from the Sun such as some Centaurs and Trans-Neptunian objects (TNOs), whereas Jupiter family comets (JFCs) such as cometary nuclei and dead comets closer to the Sun lack the reddish coloration (Luu & Jewitt 1996; Sheppard 2010). Because there is color diversity among the objects despite both Centaurs and JFCs supposedly originating in TNOs (Duncan & Levison 1997; Lowry et al. 2008), ultra-red color materials have been explained to be sublimated or destroyed as the objects approach from the Trans-Neptunian region to the inner solar system (Sheppard 2010). Accordingly, the reddish coloration and its distribution are significant for understanding the formation process of the solar system. Furthermore, it provides astrobiological insights into the origin of life (Khare et al. 1986; Cruikshank et al. 2019). The reddish color has been studied extensively in laboratory experiments and is attributed to complex refractory organic mixtures such as tholins, produced by energetic radiation on organic volatiles and their ice surfaces (Imasaka et al. 2004; Cruikshank et al. 2005; Materese et al. 2014; Materese et al. 2015). However, the tholin-type materials that have been studied are not volatile even when heated to room temperature. Due to this, explanation of the reddish color distribution is not straightforward. Although some alternative explanations for the reddish color distribution have been suggested (Jewitt 2002a; Grundy 2009; Brown et al. 2011), the color distribution is not fully elucidated and remains in debate (Dalle Ore et al. 2011).

In this study, we demonstrate the reddish coloration specific to cryogenic temperature in a laboratory experiment, that is distinct from the well-known reddish coloration of the previously reported tholins. For the coloration, cryoplasma was radiated at 85 K on methanol ($CH_3OH$) and water ($H_2O$) containing ice by feeding helium (He) with 3% nitrogen ($N_2$). Cryoplasma is non-equilibrium plasma whose gas temperature can be controlled at wide range of cryogenic temperatures below room temperature (Stauss et al. 2018), enabling coexistence of plasma with ice without melting (Sakakirbara & Terashima 2017; Sakakibara et al. 2019). This study roughly simulated energetic processes in the outer solar system via UV radiation and charged and/or excited species formed by cosmic rays and solar winds (Thompson et al. 1991; Imasaka 2004).

# 2. EXPERIMENT

## 2.1. Cryoplasma Irradiation of CH$_3$OH/H$_2$O Ice

In a custom-made cryogenic chamber equipped with a 4 K Gifford McMahon refrigerator (CKW-21, Sumitomo Heavy Industries), CH$_3$OH/H$_2$O ice grown on the top and bottom sides of the electrodes (Figure 1a) were irradiated with cryoplasma for 12 h. The setup is similar to our previous study (Sakakirbara & Terashima 2017). The top and bottom side of electrodes were composed of 150 μm-thick indium tin oxide (ITO) (10 mm in diameter)-coated glass and stainless steel (16 mm in diameter), respectively, with a gap distance of 500 μm. The CH$_3$OH/H$_2$O ice was used as a dielectric barrier to generate the cryoplasma in a dielectric barrier discharge (DBD) configuration. To prepare CH$_3$OH/H$_2$O ice, a mixture of 200 μL of water (electrical conductivity < 5 μS cm$^{-1}$, Wako pure chemical industries) and 200 μL of methanol (infinity grade, Wako pure chemical industries) was poured onto the ground electrode. The mixture was cooled at around 0.6 K/min from room temperature to 80 K by lowering the ambient temperature in the chamber. Although the initial mixture had the same volume of CH$_3$OH and H$_2$O, the ice on the top electrode might be enriched in the more volatile CH$_3$OH than H$_2$O. 3% N$_2$ gas (G1 grade) diluted with He gas (G1 grade) was pumped into the chamber at 30 sccm (standard cubic centimeters per minute) flow rate as carrier gas at $2 \times 10^3$ Pa. To generate the cryoplasma, 1.75 kV$_{pp}$ sinusoidal AC voltage was applied at 10 kHz to the top electrode with a function generator (WF1974, NF) and a high voltage amplifier (HVA4321, NF). The bottom electrode served as a ground electrode. He gas was utilized to stabilize the discharge and to offer better heat dissipation. The gas temperature of plasma was monitored by a silicon diode temperature sensor (DT470CU131.4L, Lakeshore) at 30 mm from the center of the cryoplasma, and it was maintained with PID control at 85 K with an accuracy of a few Kelvin (Sakakirbara & Terashima 2017). The cryoplasma consumed 18.5 mW, indicating that the total energy transferred to the ice was $1.2 \times 10^{-2}$ J cm$^{-2}$ s$^{-1}$. For the 12 h of plasma irradiation, this dose corresponds to between $10^4$–$10^5$ years of energetic radiation such as solar wind in the assumption of 1 keV energies with between $10^9$–$10^{10}$ particles $m^{-2}$ $s^{-1}$ (Bagenal et al. 2016). The temperature, chemicals and energetic radiation process chosen here are within the range of plausibility to the current condition of the outer solar system (Lewis 1973; Bennett et al. 2013; Dalle Ore et al. 2015).

## 2.2. TPD Analysis of the Plasma-Irradiated Ice

After the plasma irradiation of $CH_3OH/H_2O$ ice, temperature programmed desorption (TPD) experiments were carried out with a quadrupole mass spectrometer with electron impact ionization (Prisma Plus QMG 220M1, Pfeiffer Vacuum). The temperature in the chamber was increased at a linear heating rate of 1 K $min^{-1}$ from 85 K to 230 K, and the desorption of fragments from the plasma-irradiated reddish colored ice was monitored. The TPD analysis was performed up to 230 K in this study, because the melting point of $CH_3OH/H_2O$ (1:1 by volume) is approximately 230 K (Miller & Carpenter 1964). During the heating, the inner chamber was evacuated by the pumping system at $1 \times 10^{-1}$ Pa while the mass spectrometer was operated around $3 \times 10^{-5}$ Pa.

For the analysis for isotopic ice samples, 99% $^{13}CH_3OH$ (Wako pure chemical industries), 99.8% $CD_3OD$ (Wako pure chemical industries), 99.8% $D_2O$ (Wako pure chemical industries) were utilized as purchased without further purification. The isotopic ice samples were prepared by cooling $^{13}CH_3OH/H_2O$, $CD_3OD/H_2O$, or $CH_3OH/D_2O$ in 1:1 by volume from room temperature to 85 K, and then irradiated to the cryoplasma for 12 h under the same conditions as the case of $CH_3OH/H_2O$ ice sample. In all isotope-labelled experiments, the same appearance and disappearance of the reddish color was observed.

## 2.3. LC-MS and LC-MS/MS Analysis of the Residue

After the TPD analysis, the plasma-irradiated $CH_3OH/H_2O$ ice was heated to room temperature, and the residue was collected for liquid chromatography mass spectroscopy (LC-MS) and subsequent tandem mass spectroscopy (LC-MS/MS). Liquid chromatography was performed by an ultra-performance liquid chromatography system (Acquity UPLC, Waters Co., Ltd.) with a hydrophilic interaction chromatography (HILIC)-type column (Acquity UPLC BEH Amide, 2.1×150 mm, 1.7 μm, Waters Co., Ltd.). The column was kept at 40 °C. The mobile phase was composed of two components: 0.1% formic acid and acetonitrile (5:95 by volume for the first 15 min and 50:50 by volume later). 10 μL of the residue was injected as picked up and flown at a speed of 0.2 mL/min. Mass spectrometry was performed by a time-of-flight type mass spectrometer (Xevo QTof MS, Waters Co., Ltd.). Ionization of the sample was conducted by electrospray ionization at 3.0 kV capillary voltage and 30 V cone voltage. The mass spectra range was 50–1000 *m/z*. The chemical formula of fragment ions from the mass spectra was estimated by their accurate masses. For peaks with strong intensity, subsequent tandem mass spectroscopy was conducted. Chemical structures in the mass spectra were estimated from both their accurate masses and the mass spectra of their peaks.

# 3.  RESULTS

During the irradiation of the cryoplasma, photographs of ice were taken from the upper side of the electrode for color monitoring, and the optical emission from the cryoplasma was observed through the top electrode coated with transparent ITO (Figure 1a). We found that reddish coloration appeared on the ice and became more prominent with longer plasma irradiation (Figure 1b). On the other hand, the intensities of the optical emission lines from the cryoplasma decreased with time in UV and visible range (Figure 1c). In other words, transmittance of the ice at UV and visible wavelengths decreased (Figure 1c), assuming constant emission intensity of the cryoplasma during the irradiation. The assumption is reasonable as the power consumption of cryoplasma during the irradiation was nearly constant (Figure S2). Therefore, the reddish color was attributed to the absorption by the product synthesized through the plasma irradiation. When using cryoplasma without $N_2$ gas flow as a control experiment, the reddish color was not observed visually (Figure S3). Thus, reactive nitrogen species from plasma, such as those identified in the emission spectrum of $N_2^+$ ions and excited $N_2$ molecules (shown in Figure 1c), should play a key role in the appearance of the reddish color.

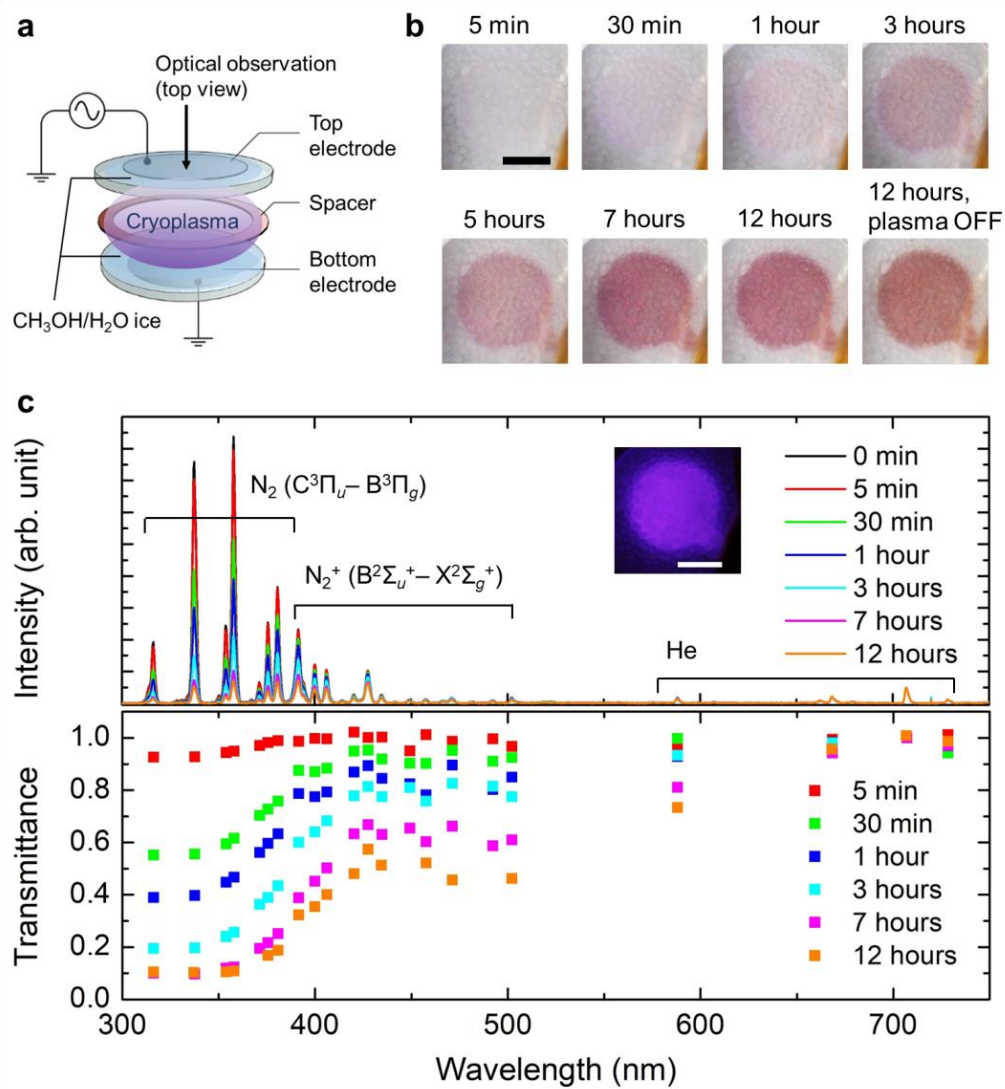

**Figure 1. Reddish coloration of the CH₃OH/H₂O ice during plasma irradiation.** (**a**) Schematic illustration of the electrodes of the cryoplasma. (**b**) Photographs showing the changes in the reddish coloration of the ice during plasma irradiation. The reddish color was not the color of the plasma itself, as shown in the photograph taken after the plasma was turned off. The color was not degraded even after three days. The scale bar is 5 mm. (**c**) Optical emission spectra of the plasma (top panel) and transmittance of the ice (bottom panel) at different plasma irradiation durations. The transmittance at a given wavelength was calculated using the rate of decrease of the plasma emission intensity. The inserted photograph shows the top view of the cryoplasma. The scale bar is 5 mm.

To further analyze the reddish coloration, TPD experiments were performed. The post plasma-irradiated ice was heated at 1 K min$^{-1}$ from 85 K to 230 K at $1\times10^{-1}$ Pa. The reddish color started fading at 120 K and subsequently disappeared at 150 K, as shown in Figure 2a. The reddish color was not reproduced when the temperature was conversely lowered from 165 K to 85 K. It was still not reproduced when the same ice sample was subjected to 12 h of the plasma irradiation at 175 K (Figure S4). In contrast, the reddish color was sustained when cooling the post plasma-irradiated ice from 85 K to 20 K (Figure S5). Therefore, the reddish color was specific to cryogenic temperatures. Since the disappearance of the reddish color was not reversible with respect to temperature, it might be due to the desorption of reddish substances and/or an irreversible chemical change of the reddish substance into colorless materials.

A quadrupole mass spectrometer was used to measure the desorption from the post plasma-irradiated ice simultaneously with the reddish color disappearance. An increase in the intensity of the mass spectra signals was detected at some mass-to-charge ratio ($m/z$) in accordance with the disappearance of the reddish color. In particular, $m/z = 30, 46, 60, 61, 72$ and 73 showed noticeable increases, as shown in the TPD spectra in Figure 2b. These $m/z$ signals could correspond to fragments of desorbed reddish materials or by-products derived from the transformation of reddish to colorless materials. To further investigate the desorption of substances that accompanied the color disappearance, TPD analysis for isotopic ice samples was also conducted. $^{13}CH_3OH/H_2O$ ice, $CD_3OD/H_2O$ ice, and $CH_3OH/D_2O$ ice were irradiated with the cryoplasma under the same conditions as described above. The $m/z$ shifts between labelled and unlabeled methanol- and water-containing ice allowed estimation of the number of C or H atoms in the monitored substances.

Initially, signals at $m/z = 60$ and 61 showed clear shifts to higher $m/z$ when methanol or water ices were isotopically labelled (right panel of Figure 2c). Carbon-13 labelling of methanol revealed that substances at $m/z = 60$ and 61 contained one methanol-derived C atom, while D labelling of methanol revealed two or four methanol-derived H atoms for $m/z = 60$ and three methanol-derived H atoms for $m/z = 61$. D labelling of water ice showed that both $m/z = 60, 61$ contained no water-derived H atom. Some possible molecular formulas of the substances that satisfy the isotopic analysis results are $CH_2NO_2$ or $CH_4N_2O$ for $m/z = 60$ and $CH_3NO_2$ for $m/z = 61$. Both of these are nitrogen-containing unsaturated compounds that could possess C=N, N=O or amide bonds, as shown in the possible chemical structures in Figure S6. Substances at $m/z = 72$ and 73 could possess unsaturated groups and/or nitrogen as well, as shown in Figure S6–S8.

This assignment of the fragments was consistent with the TPD and isotopic analysis for $m/z = 30$ and 46. The spectra at $m/z = 30$ and 46 did not show any shift in the peak positions when any of the isotopic labelled ice was used (left panel of Figure 2c, Figure S7). This implies that fragments at $m/z = 30$ and 46 do not contain any C or H atoms derived from methanol and water ice. Therefore, peaks at $m/z = 30$ and 46 likely stem from molecules composed entirely of N and O atoms, such as nitric oxide (NO) and nitrogen dioxide ($NO_2$), respectively. However, NO is colorless, and $NO_2$ converts to colorless dinitrogen tetroxide ($N_2O_4$) at lower temperatures below the melting point of $N_2O_4$ (261.9 K), although $NO_2$ appears to be reddish in color. Although $NO_2$ might be formed directly at cryogenic temperatures, the cryogenic presence of $NO_2$ is unclear because no data of $NO_2$ vapor pressure has been proposed (Fray & Schmitt 2009). Thus, tentatively, a plausible situation is these nitrogen and oxygen containing compounds are fragments from our reddish materials.

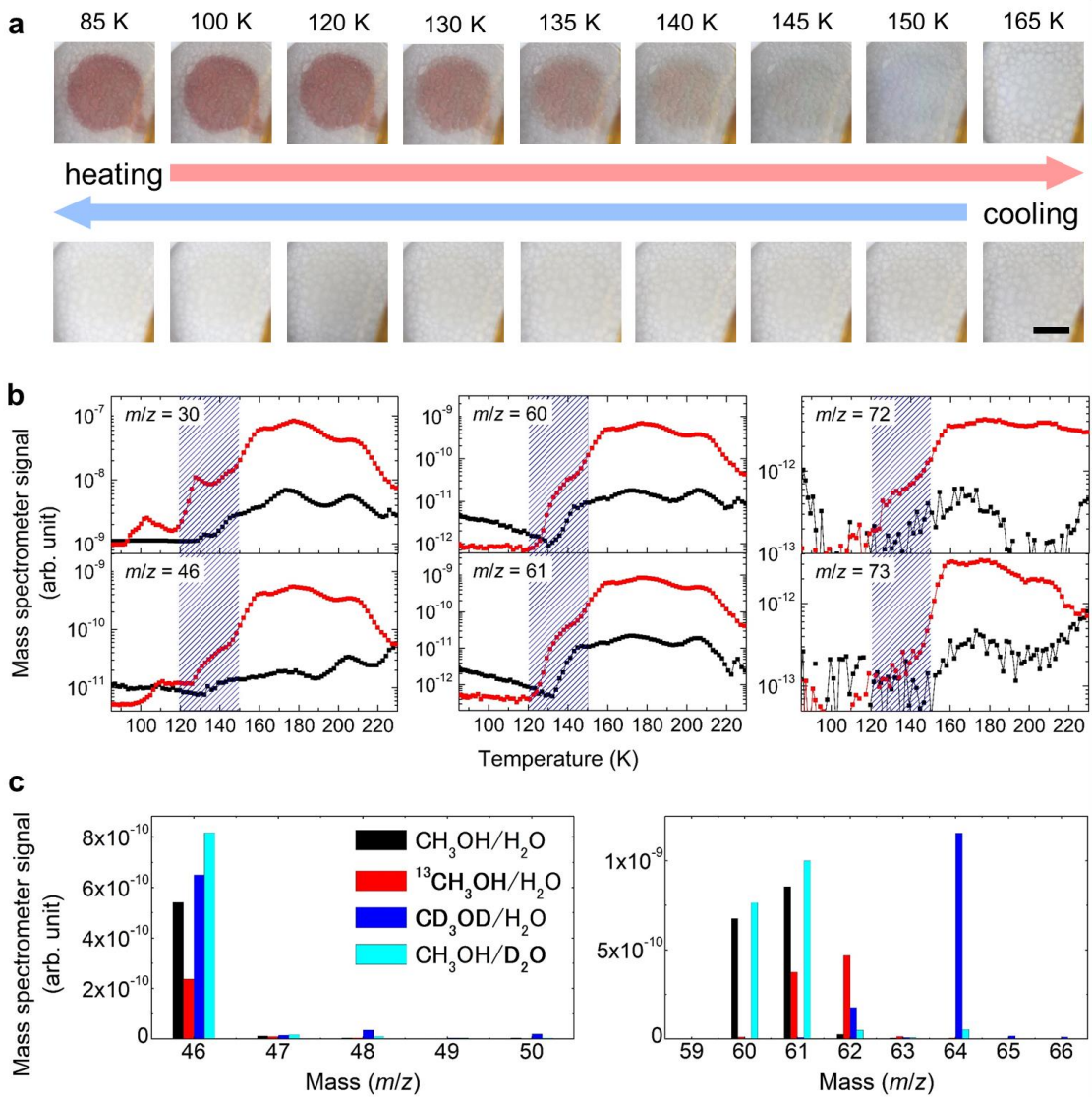

**Figure 2. Heating of the post plasma-irradiated CH₃OH/H₂O ice.** (**a**) Photographs showing the disappearance of the reddish color during the heating. The scale bar is 5 mm. (**b**) TPD spectra indicating desorption of fragments from the post plasma-irradiated ice. The red and black lines represent plasma irradiation with and without $N_2$ gas input, respectively. The blue shaded regions indicate the temperature range where the reddish color was seen to be disappearing. (**c**) Mass spectra of the isotope-labeled analysis of the post plasma-irradiated ice. Mass spectra at 180 K are shown here. The black, red, blue, and cyan indicators represent the post plasma-irradiated ice of CH₃OH/H₂O, ¹³CH₃OH/H₂O, CD₃OD/H₂O, and CH₃OH/D₂O, respectively.

The analysis of the colorless liquid residue obtained at room temperature supports the idea that nitrogen-containing and unsaturated organic compounds could have caused the reddish coloration. After heating the post plasma-irradiated $CH_3OH/H_2O$ ice to room temperature, the residue was analyzed by LC-MS and further mass spectroscopy (LC-MS/MS). Synthesis of various organic compounds with larger masses than the original $CH_3OH/H_2O$ ice, such as glycols, carboxylic acids, amines and amides, were identified as shown in Figure 3 and Figure S10. Some of these compounds contain C=C and C=N bonds that absorb UV and visible radiation in the conjugated system, thus giving rise to a colored appearance. In terms of astrobiology, such structures and functional groups are essential building blocks of prebiotic organic molecules such as proteins (Berstein et al. 2002; Munoz Caro et al. 2002) and genetic materials (Meinert et al. 2016). This implies that the observed reddish color is closely related to the possible existence of prebiotic organic compounds (Cruikshank et al. 2019), and especially prebiotic substances that can be stable only at cryogenic environments in the outer solar system.

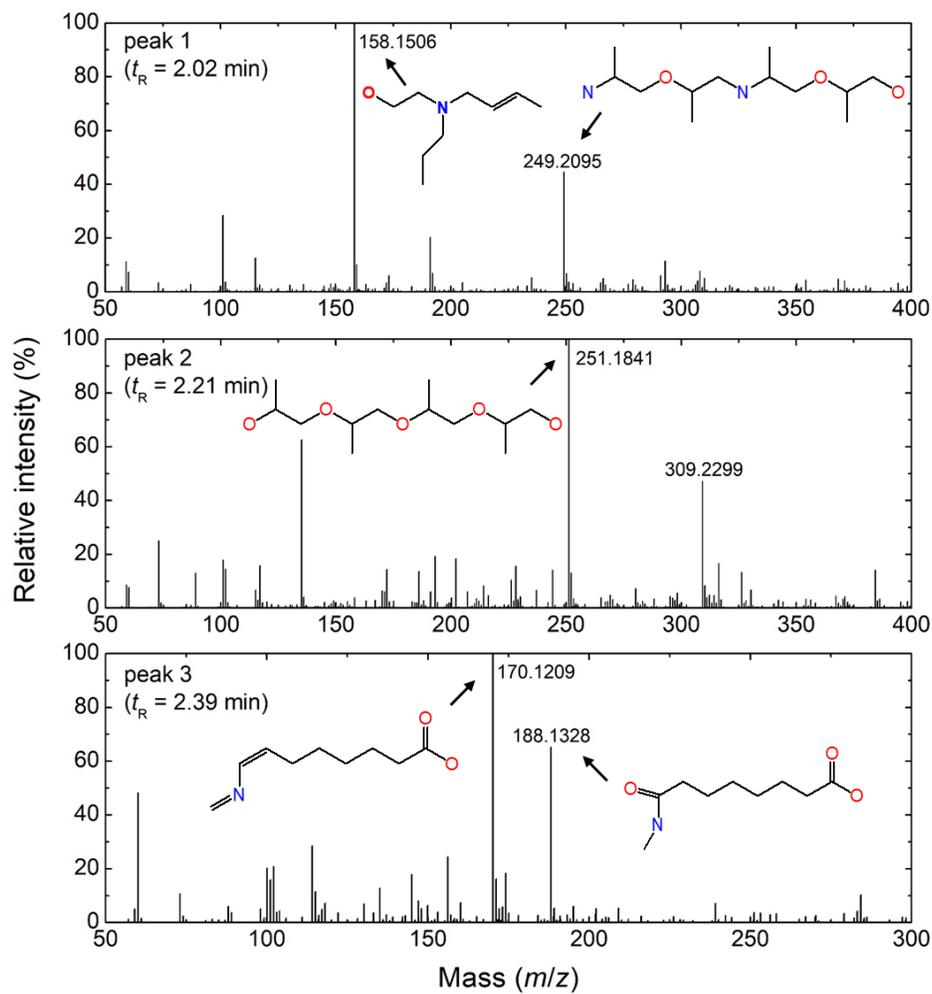

**Figure 3. LC-MS spectra of the residue of the plasma-irradiated CH₃OH/H₂O ice at room temperature.** Mass spectra corresponding to different retention times ($t_R$) of total ion chromatogram are shown in Figure S9. The chemical structures of representative peaks identified by MS/MS analysis (Figure S10) are also depicted.

# 4. DISCUSSION

In the outer solar system, reddish colors have been explained by complex organic tholins, which contain delocalized electrons in the conjugated unsaturated bonds of aliphatic and aromatic hydrocarbons with some amounts of substituted nitrogen (Imasaka et al. 2004; Cruikshank et al. 2005). Excitation of delocalized electrons is accompanied by absorption at UV and visible wavelengths, which correspond to the reddish color. The reddish materials in this study can also be attributed to complex organic materials, considering the absorption at UV and visible wavelengths (Figure 1c) and the nitrogen-containing unsaturated fragments (Figure 2b, 2c). However, the temperature dependent behavior in lower temperature in this study has not been observed in previously reported reddish coloration of refractory tholins. This might imply a significant contribution of smaller and more volatile colorants to the appearance of cold surfaces in the outer solar system.

The color diversity of icy objects in the outer solar system, and especially the absence of ultra-red coloration closer to the Sun, has been previously explained by the variation of compositions depending on the body environment (Jewitt 2002a; Grundy 2009; Sheppard 2010; Brown et al. 2011), and considering reddish organic tholins and colorless volatile species (Cruikshank et al. 2005). However, the explanation is not straightforward and still debatable because the well-known organic tholins are non-volatile even when warming to room temperature. Here, the formation of the cryogenic-specific reddish coloration could offer a new possible explanation to the reddish color distribution in the outer solar system. The cryogenic-specific reddish materials disappear by sublimation or conversion to other colorless compounds at warmer cryogenic temperatures, as an icy body travels from the Trans-Neptunian region to the inner solar system, such as Centaurs and thence Jupiter family region. This scenario implies that the body coloration and its diversity might be able to serve as a probe of current temperature and/or temperature history. The concept of reddish materials that are stable only at cryogenic temperatures is potentially useful for further investigation of color diversity observed in the outer solar system, including many reddish objects, and would contribute to the further understanding of the nature and formation of the solar system.

# 5. CONCLUSION

In this study, we demonstrated reddish coloration that is stable only at cryogenic temperatures by irradiating methanol- and water-containing ice with nitrogen-containing cryoplasma at 85 K. The reddish color visually faded and disappeared at 120–150 K as the ice was heated, unlike well-known refractory organic tholins that are stable even when heated to room temperature. By TPD analysis with isotopically labelled ices and LC-MS/MS analysis of the residue, the reddish material was suggested to be nitrogen-containing organic compounds. The temperature dependence of reddish coloration under cryogenic conditions could provide a new possible explanation for the absence of ultra-red coloration closer to the Sun in the outer solar system. Our result implies that a reddish material specific to cryogenic environments could be useful for the investigation of color diversity and formation mechanism of the outer solar system. Moreover, it is worthwhile to mention that cryoplasma was demonstrated to be a novel technology for accelerating the investigation of chemistry and materials science in cryogenic space environments including the outer solar system.

## ACKNOWLEDGEMENTS

This work was partially supported by a Grant-in Aid for Scientific Research (A) (Grant No. 24246120) from the Ministry of Education, Culture, Sports, Science and Technology of Japan, and a Grant-in Aid for JSPS Fellows (Grant No. 19J13045). One of the authors (N.S.) was supported by a Grant-in-Aid via a Japan Society for the Promotion of Science (JSPS) Research Fellowship. We thank Foundation for Promotion of Material Science and Technology of Japan (MST) for the technical support of the LC-MS/MS analysis. N.S. thank S. Morita and C. Ode for fruitful discussions.

Supplementary information for

# Cryogenic specific reddish coloration by cryoplasma: New explanation to color diversity of outer solar system objects

Noritaka Sakakibara*, Phua Yu Yu, Tsuyohito Ito, Kazuo Terashima*

Department of Advanced Materials Science, Graduate School of Frontier Sciences, The University of Tokyo, 5-1-5 Kashiwanoha, Kashiwa, Chiba 277-8561, Japan.

*Corresponding author. E-mail: n.sakakibara@plasma.k.u-tokyo.ac.jp, kazuo@plasma.k.u-tokyo.ac.jp

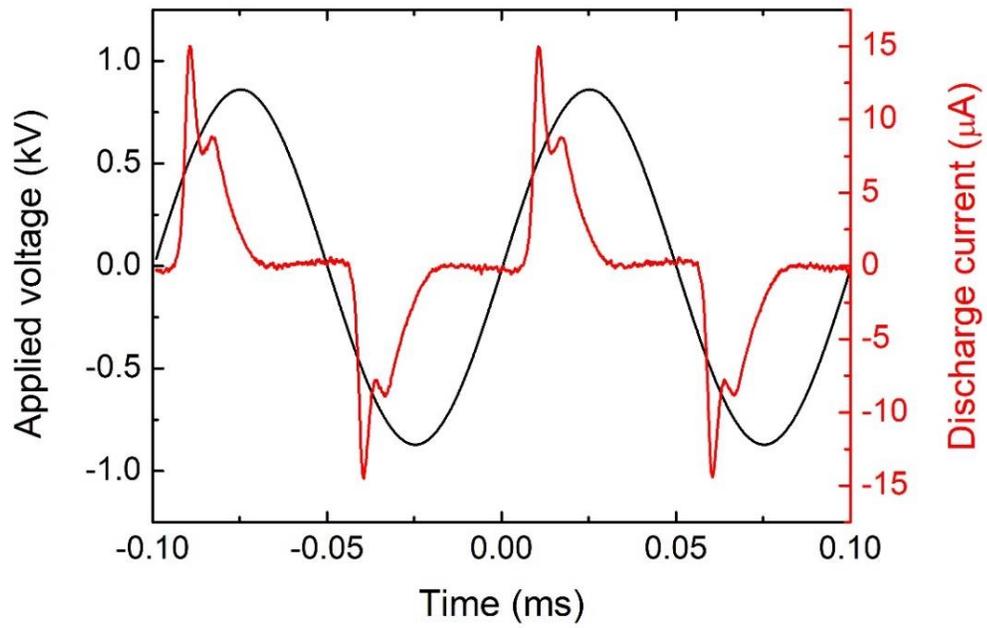

**Figure S1.**
*I-V* characteristics of the cryoplsama. The black and red curves indicate the applied voltage to the electrode and the discharge current of the cryoplasma, respectively.

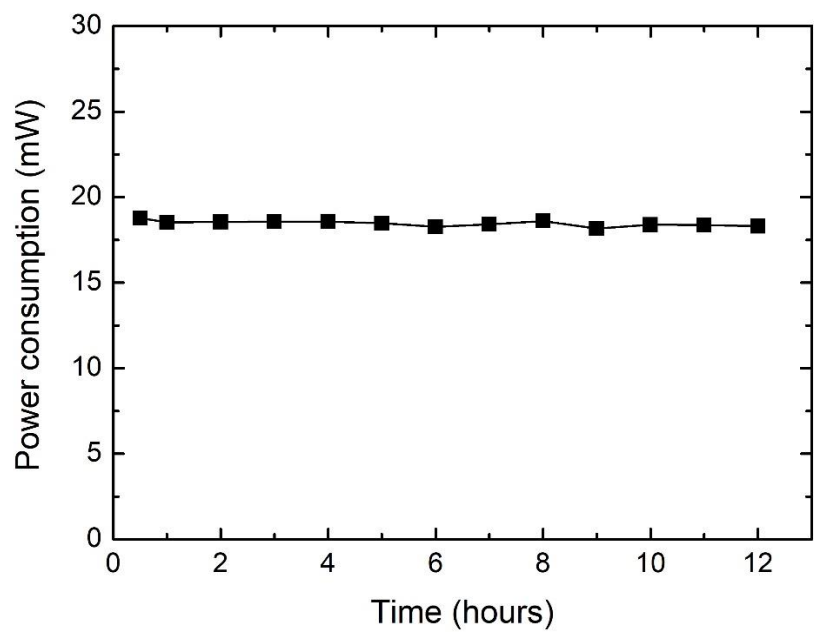

**Figure S2.**
Temporal change in power consumption of the cryoplasma. The power consumption was stable within ±0.4 mW. Therefore, the emission intensity of the cryoplasma during the 12 hours of irradiation was considered to be constant.

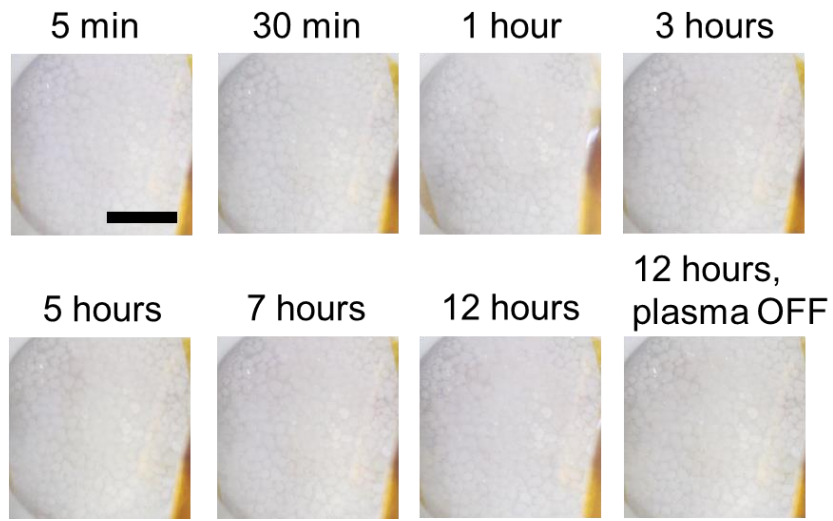

**Figure S3.**
Photographs of the ice during the cryoplasma irradiation without nitrogen gas flow at 85 K. Photograph of the post-irradiation ice with the plasma turned off after 12 hours of irradiation is also illustrated. The scale bar is 5 mm. The plasma-irradiated ice did not show any visible color change when nitrogen gas was not introduced.

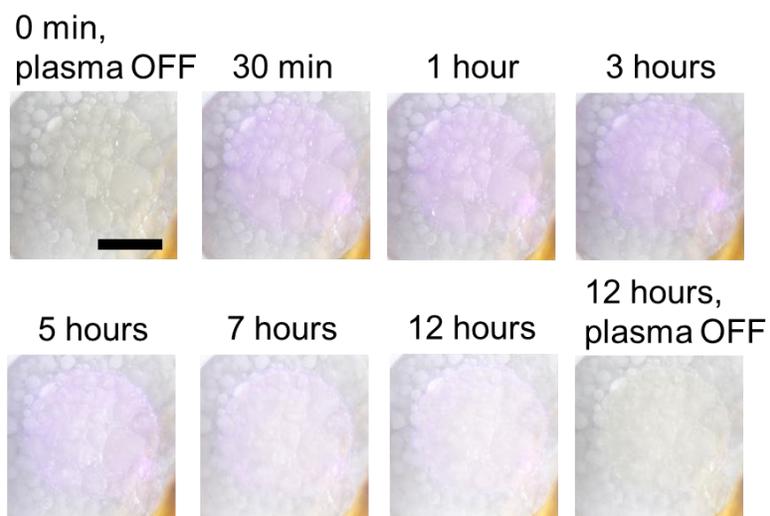

**Figure S4.**
Photographs of the ice during the cryoplasma irradiation at 170 K. Photographs of the pre-irradiation and post-irradiation are also shown. The scale bar is 5 mm. The plasma-irradiated ice did not show any visible color change when irradiated at 170 K.

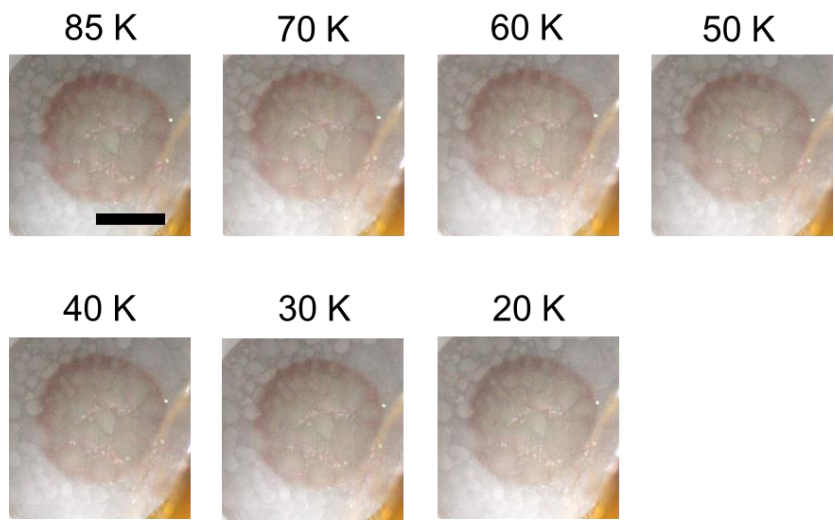

**Figure S5.**
Cooling of the post plasma-irradiated $CH_3OH/H_2O$ ice. Photographs show that the reddish color was visibly maintained throughout the cooling to 20 K. The scale bar is 5 mm. The non-uniform reddish color was attributed to the non-uniform plasma generation at this time. The plasma was produced more strongly near the edge of the top electrode during this experiment.

**m/z = 60**

CH$_2$NO$_2$ (1.5) 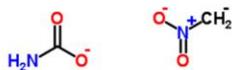

CH$_4$N$_2$O (1) 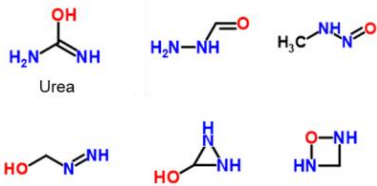

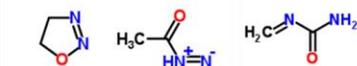

**m/z = 72**

C$_2$H$_4$N$_2$O (2) 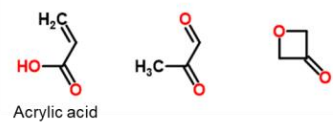

C$_3$H$_4$O$_2$ (2)

CH$_4$N$_4$ (1) 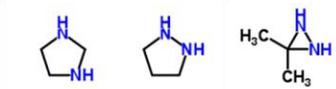

C$_3$H$_8$N$_2$ (1)

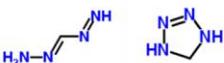

C$_4$H$_8$O (1) 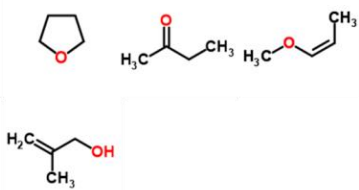

**m/z = 61**

CH$_3$O$_2$N (1) 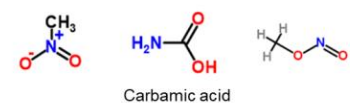

**m/z = 73**

CHN$_3$O (3) 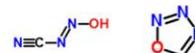

C$_2$H$_3$NO$_2$ (2) 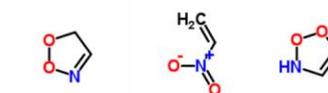

C$_2$H$_7$N$_3$ (1) 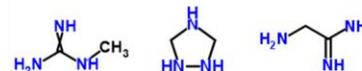

C$_3$H$_9$NO (1) 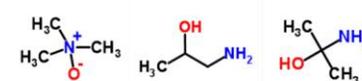

C$_4$H$_{11}$N (1) 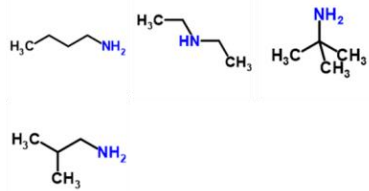

**Figure S6.**
Examples of various molecular formula satisfying the molecular masses. For *m/z* = 60 and 61, the molecular formula suggested above are also in good agreement with the results of isotopic analyses. The chemical structures are cited from a free chemical structure database ChemSpider (www.chemspider.com/). Numbers inside the parentheses indicate the degree of unsaturation.

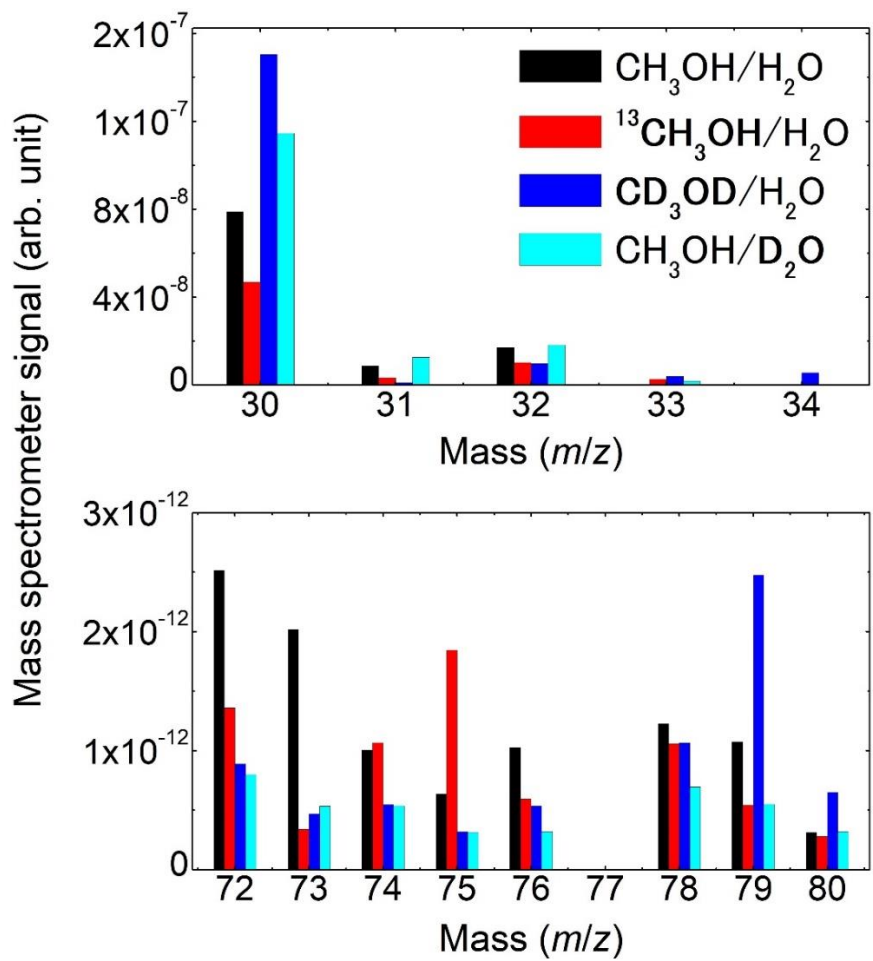

**Figure S7.**

Mass spectra of the isotope-labeled analysis of the post plasma-irradiated ice at different *m/z* ranges extended from the spectra in Figure 2C. Mass spectra at 180 K are shown here. Isotopic analyses at *m/z* = 30 (top panel) showed similar shift in the peak positions as *m/z* = 46 (Figure 2C). Isotopic analyses at *m/z* = 72 and 73 (bottom panel) did not show clear shifts in the mass spectra peaks because fragments at higher masses (*m/z* = 74–80) also desorbed during the disappearance of the color (Figure S8), and their mass spectra signals increased. Various molecular formulas are possible for *m/z* = 72 and 73 as shown in Figure S6, some of which are unsaturated and/or nitrogen containing.

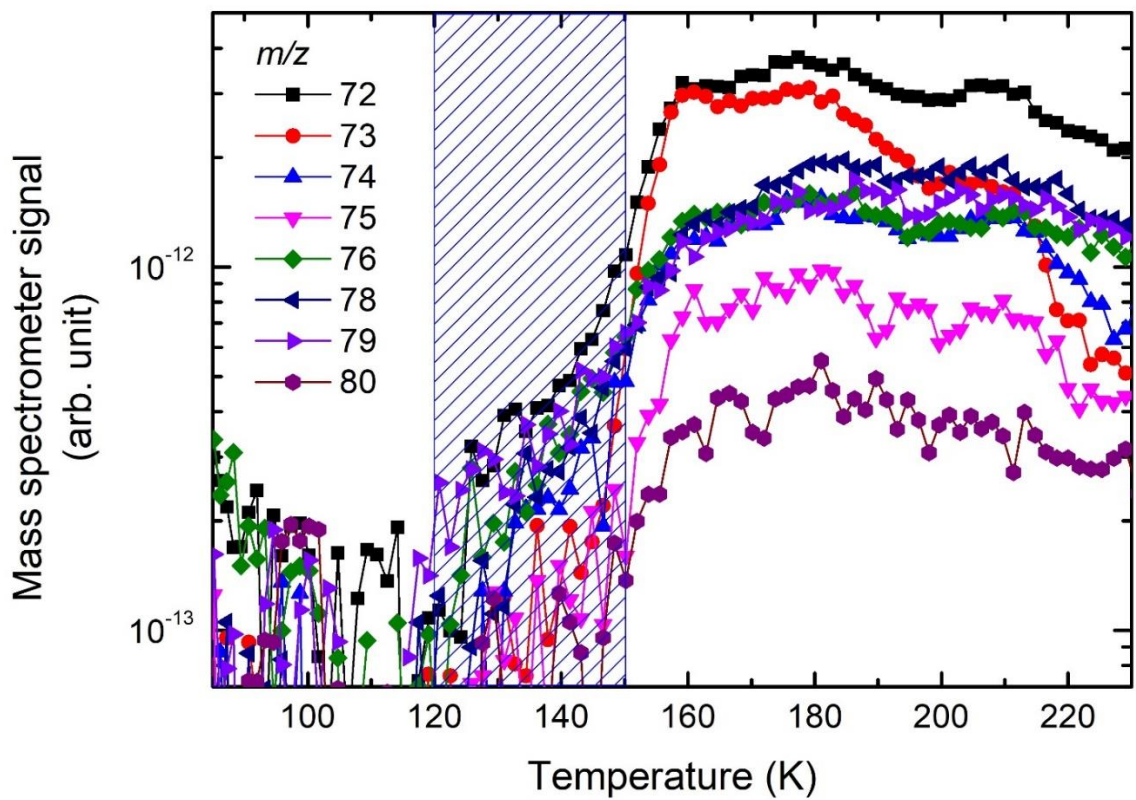

**Figure S8.**
TPD spectra of the post plasma-irradiated $H_2O/CH_3OH$ ice for $m/z = 72–80$. The spectra for $m/z = 77$ could not be monitored correctly because of an inevitable noise from the QMS equipment.

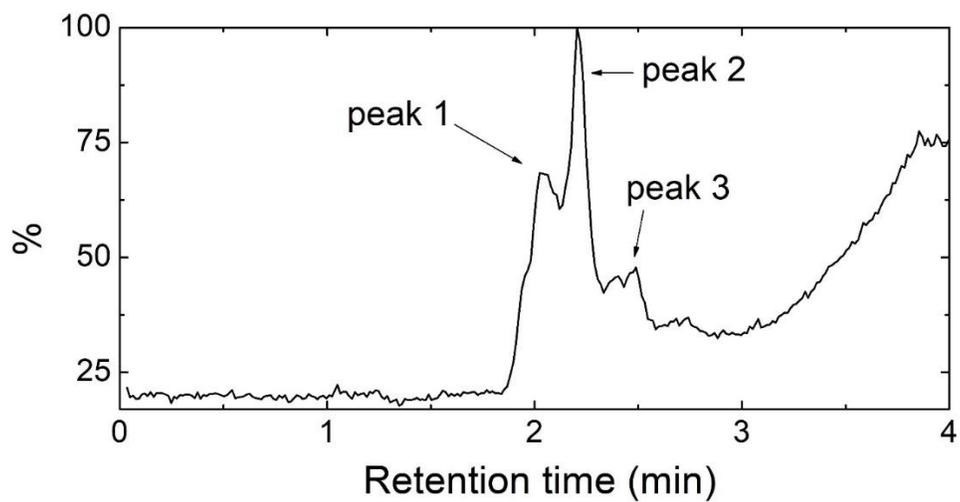

**Figure S9.**

Total ion chromatogram by the liquid chromatography of the sample residue at room temperature. Further mass spectroscopy was performed for the peaks of different retention times ($t_R$), i.e., peak 1 at $t_R$ = 2.02 min, peak 2 at $t_R$ = 2.21 min, and peak 3 at $t_R$ = 2.39 min.

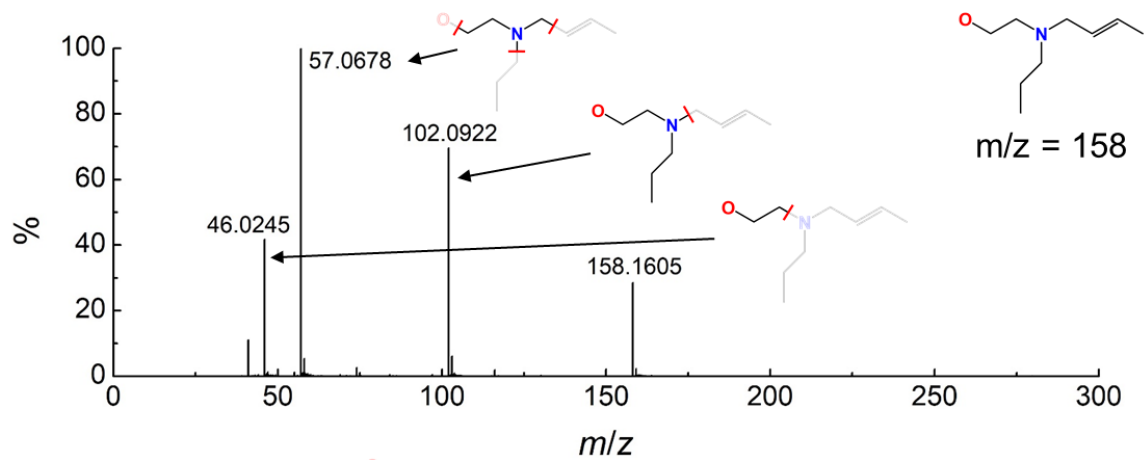

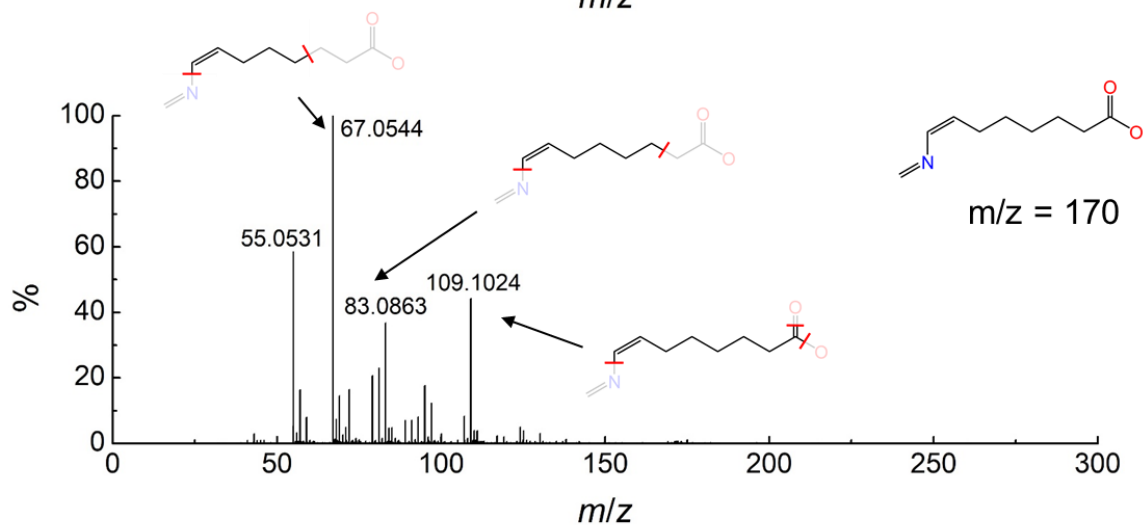

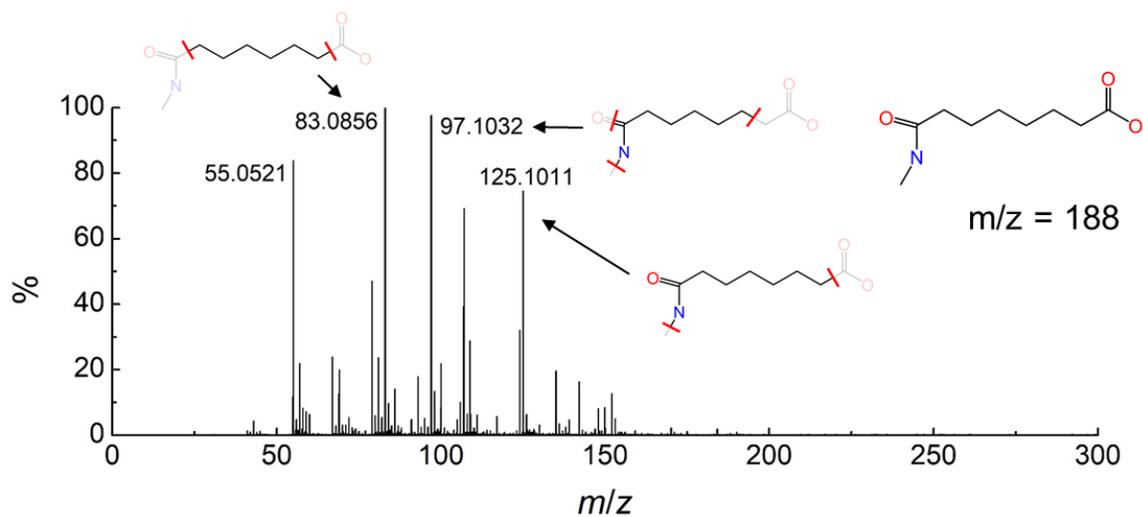

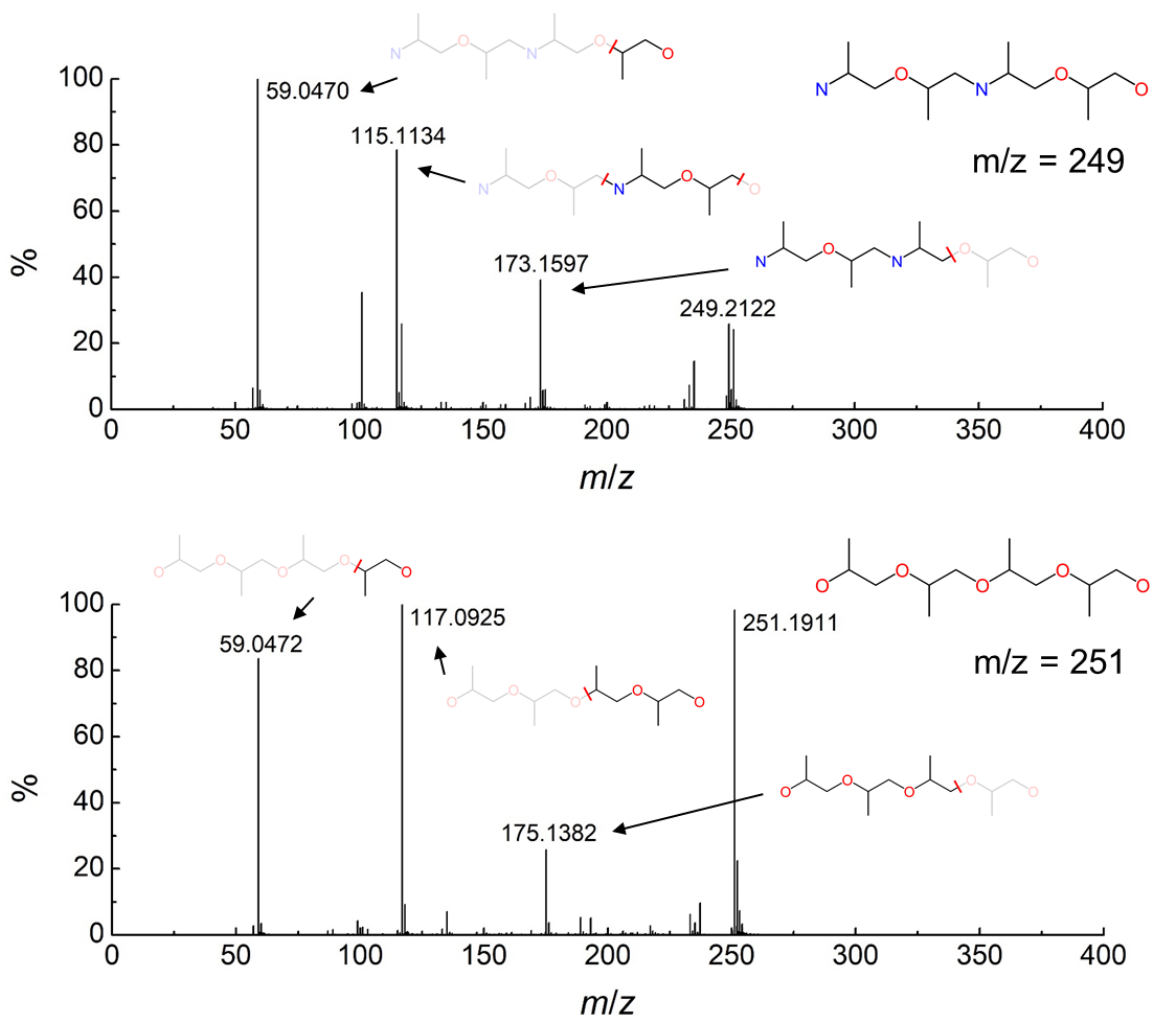

**Figure S10.**

LC-MS/MS spectra of the peaks at $m/z$ = 158, 170, 188, 249, and 251 in the LC-MS spectra in Figure 3. MS/MS analysis was carried out to estimate chemical structures of the peaks by monitoring the fragments from the substance of the target $m/z$ peak. The chemical structures of fragments estimated for the observed MS-MS spectra peaks are also depicted.